# Fractional matter coupled to the emergent gauge field in a quantum spin ice

**Victor Porée[1†], Han Yan,[2†] Félix Desrochers,[3†] Sylvain Petit[4‡], Elsa Lhotel[5‡], Markus Appel[6], Jacques Ollivier[6], Yong Baek Kim,[3] Andriy H. Nevidomskyy[2] & Romain Sibille[1*]**

[1]Laboratory for Neutron Scattering and Imaging, Paul Scherrer Institut, 5232 Villigen PSI, Switzerland, [2]Department of Physics & Astronomy, Rice University; Houston, TX 77005, USA, [3]Department of Physics, University of Toronto, Toronto, Ontario M5S 1A7, Canada, [4]LLB, CEA, CNRS, Université Paris-Saclay, CEA Saclay, 91191 Gif-sur-Yvette, France, [5]Institut Néel, CNRS – Université Grenoble Alpes, 38042 Grenoble, France, [6]Institut Laue-Langevin, CS 20156, F-38042 Grenoble Cedex 9, France, [*]email: romain.sibille@psi.ch [†]These authors contributed equally to this work. [‡]These authors contributed equally to this work.

**Electronic spins can form long-range entangled phases of condensed matter named quantum spin liquids[1–4]. Their existence is conceptualized in models of two- or three-dimensional frustrated magnets that evade symmetry-breaking order down to zero temperature. Quantum spin ice (QSI) is a theoretically well-established example described by an emergent quantum electrodynamics, with excitations behaving like photon and matter quasiparticles[5–6]. The latter are fractionally charged and equivalent to the 'spinons' emerging from coherent phases of singlets in one dimension, where clear experimental proofs of fractionalization exist[7–9]. However, in frustrated magnets it remains difficult to establish consensual evidence for quantum spin liquid ground states and their fractional excitations. Here, we use backscattering neutron spectroscopy[10] to achieve extremely high resolution of the time-dependent magnetic response of the candidate QSI material $Ce_2Sn_2O_7$** (refs. 11,12)**. We find a gapped spectrum featuring a threshold and peaks that match theories[13–19] for pair production and propagation of fractional matter excitations (spinons) strongly coupled to a background gauge field. The multiple peaks are a specific signature of the $\pi$-flux phase of QSI[16-19], providing spectroscopic evidence for fractionalization in a three-dimensional quantum spin liquid.**

The idea that certain phases of condensed matter have "quantum orders" alludes to the description of their electronic correlations with an effective low-energy gauge theory, without spontaneous symmetry breaking[20–21]. The emergent gauge field reflects the long-range and/or topological entanglement of a complex ground-state wavefunction, which results in a variety of exotic properties such as excitations carrying fractional quantum numbers. A famed example is the collective behavior of a two-dimensional electron gas where electrons acquire a fractional elementary charge[22]. Similar states termed quantum spin liquids (QSL) are predicted to emerge in models of two- and three-dimensional frustrated magnets[1–4]. Their effective low-energy description is a deconfined gauge theory where quasiparticles that carry spin 1/2 and no charge, known as spinons, can propagate coherently with the background gauge field. However, because the fractional spin excitations interact strongly with the background gauge field under which they are charged, their dynamics is highly non-trivial. The symmetries of the underlying crystal structure can additionally enrich topological phases: spinons can carry fractional crystal momentum, leading to enhanced periodicity of the excitation spectrum in momentum[23-24,19].

A prototypical model of a three-dimensional frustrated magnet is the spin ice[25], whose magnetic degrees of freedom reside on a lattice of corner-sharing tetrahedra where each motif results in a local '2-in-2-out' constraint. The classical limit of this model is called classical spin ice (CSI) and consists of a macroscopically degenerate manifold of ground states obeying this local rule reminiscent of the arrangement of hydrogens in water ice[26] (Fig. 1**a**). Such physics is realized in rare-earth pyrochlore materials with large uniaxial magnetic moments, where thermally-driven spin flips create pairs of emergent fractional quasiparticles called "magnetic monopoles" (Fig. 1**b-c**)[27]. These quasiparticles interact through an effective Coulomb potential, which, in materials like $Ho_2Ti_2O_7$ (ref. 28), arises from classical dipole–dipole forces. It is theoretically well established that a true QSL can be stabilized in rare-earth pyrochlores with an isolated ground-state doublet (an effective spin-1/2) by nearest-neighbor transverse interactions $J_\pm$ acting perturbatively on CSI states[5–6,29–31]: $\mathcal{H}_{QSI} = \mathcal{H}_{CSI} + \mathcal{H}_\pm = \sum_{(i,j)} J_\parallel S_i^\alpha S_j^\alpha - J_\pm (S_i^+ S_j^- + S_i^- S_j^+)$ where $\alpha$ designates the pseudo-spin components forming the ice manifold under the influence of the dominant nearest-neighbor interaction $J_{//}$. The dominant tunneling process of this quantum spin ice (QSI) is a ring exchange term ($J_{ring} \sim J_\pm^3/J_{//}^2$) that corresponds to flipping loops of head-to-tail spins on a hexagonal plaquette[5] (Fig. 1**d**). The ring exchange terms have local symmetry properties – a U(1) invariance – making their effective lattice gauge theory analogous to quantum electrodynamics (QED). The sign of the transverse interaction translates into distinct QSI phases where the hexagonal plaquettes are threaded by static 0 ($J_\pm > 0$) and $\pi$ ($J_\pm < 0$) fluxes of the emergent gauge field[32-33,19]. At temperatures $T \approx J_{ring}$, the QSI ground state is characterized by gapless, linearly dispersing excitations, which are transverse fluctuations of the gauge field and correspond to the photons of the emergent QED. At higher temperatures $J_{ring} \ll T \ll J_{//}$, however, thermal fluctuations destroy part of the quantum coherence and gradually restore a CSI[29–31]. The exotic nature of QSI also stands out from its gapped, fractional excitations – spinons, which are characterized by a larger energy scale set by $J_{//}$. They correspond to "magnetic monopoles" (electric charges in QED language) executing coherent quantum motion[5–6].

Neutrons can create spin-flip excitations leading to integer changes of the total spin, which in a QSL is expected to generate pairs of spinons that separate and execute quantum motion under the constraints of the emergent background gauge field. Here, we present neutron spectroscopy data of a candidate QSI material – $Ce_2Sn_2O_7$ (refs. 11–12), providing a wavevector-integrated spectrum of its magnetic response with μeV resolution. From a technical perspective, our findings demonstrate an advance in terms of energy resolution improved by more than an order of magnitude compared to other studies, allowing quantitative comparisons with theories for spinon dynamics in QSI.

We first present inelastic neutron scattering (INS) data acquired using a time-of-flight (TOF) spectrometer (Fig. 2**a**), at different temperatures across the dominant energy scale in $Ce_2Sn_2O_7$ ($J_{//} \approx 50$ μeV $\approx 0.6$ K)[12]. The magnetic response is essentially inelastic, as shown by the lack of temperature dependence of the elastic line (Fig. 2**c**). We use the highest temperature spectrum measured at 5 K, which is well above the correlated regime in this material, to evaluate the magnetic scattering $S(E)$ at lower temperatures by difference. Fig. 2**b** shows the imaginary part of the dynamic spin susceptibility calculated as $\chi''(E) = S(E) \times [1 - \exp(-E/k_B T)]$, where $E$ is the neutron energy transfer and $k_B$ is the Boltzmann's constant. This data shows the typical magnetic response in cerium pyrochlores[12,34–35]: a continuum of spin excitations, as expected from spinons[13–17], peaked close to the energy of the dominant exchange interaction $J_{//}$. We fit these spectra using a phenomenological Lorentzian peak shape to capture their temperature evolution (Fig. 2**d**-**f**). The center of the band is temperature independent within the resolution of the measurement, and the intensity of the continuum increases while its width reduces upon cooling. This evolution occurs mostly below 1 K, which is consistent with changes previously reported in bulk magnetic susceptibility and diffuse magnetic scattering[11–12]. At the lowest measured temperature $T \sim 0.2$ K, the data suggest a gapped spectrum with a non-trivial density of states (DOS) as shown in the inset of Fig. 2**b**. The TOF energy resolution of about 11 μeV, however, does not allow to characterize the DOS in sufficient details.

While in TOF data the energy resolution is largely determined by the value and spread of the incident neutron wavelength, in backscattering spectroscopy it is mainly limited by the properties of crystal analyzers (Fig. 3**a**)[10]. Fig. 3**b** presents data acquired using a typical backscattering geometry where the incident energy is varied by Doppler effect, covering a window of ±30 μeV around the elastic line with a high resolution (HR) of 0.7 μeV. This allows to observe the gap expected for spinons in a QSI (Fig. 3**b** inset). We also performed another experiment on the same instrument but using a recently developed option of 'backscattering and time-of-flight spectrometer' (BATS)[36]. The latter provides a larger energy window of ±250 μeV, covering the entire bandwidth of the continuum in $Ce_2Sn_2O_7$. The increase in energy range comes at the cost of a coarser resolution of 3.3 μeV in our data, which still provides a more than threefold improvement in resolution over the TOF data in Fig. 2, and allows to capture fine details throughout the entire continuum.

In Fig. 3**c** we show the dynamic spin susceptibility $\chi''(E)$ obtained from combined HR and BATS data. At the base temperature of these experiments ($T \approx 0.17$ K), $\chi''(E)$ can be well fitted using three Gaussian peaks of unconstrained widths. The maximum of the spectrum is reproduced by the main Gaussian peak located around 50 μeV, above which a gradual intensity decrease is observed, well accounted for by the two weaker Gaussian peaks around 100 μeV and 150 μeV, resulting in an overall asymmetric spectrum. Comparing the total fitted curve with an extrapolation of the experimental data points shows that the latter deviates slightly from the former at the lowest energy transfers, suggesting a threshold behavior at the bottom edge of the gapped continuum. At a higher temperature close to the uncorrelated regime ($T \approx 0.8$ K), $\chi''(E)$ shows much weaker inelastic scattering, in excellent agreement with the TOF data.

A continuous spectrum of excitations is usually taken as a hallmark of QSL states, but there can be alternative explanations for their existence, such as disorder[37]. These continua are therefore less deterministic of fractional quasiparticles than, for instance, jumps in the electric conductance of a two-dimensional electron gas[22]. However, using combinations of analytical and numerical methods applied to the case of QSI[14–18,38], theory has recently focused on studying the DOS of spinons. Importantly, these predictions provide more specific features than just a continuum, highlighting the structure of the underlying gauge field theory. We provide an exhaustive comparison with the available theoretical models and discuss their applicability and limitations.

We first consider analytical results for the quantum dynamics of spinons hopping on a CSI background[14], which is relevant at finite temperatures $J_{ring} \ll T \ll J_{//}$. The spinon hopping is constrained by the flippable spins in the CSI background – a condition that makes its propagation deviate significantly from a free hopping model and results in the unique threshold and asymmetry in the wavevector-integrated DOS[14]. This model captures the gross features observed in the excitation spectrum of $Ce_2Sn_2O_7$ (solid blue line in Fig. 4**a**). The threshold and asymmetry of the continuum are important experimental observations, since they reflect the effect of the background gauge field on the dynamics of the fractional quasiparticles[14]. The fitted exchange parameters based on the analytical hopping model, $J_{//} = 48$ μeV and $J_{\pm} = -5.2$ μeV, are in good agreement with previous estimates based on fits of bulk thermodynamic properties at the mean-field level[12]. In the context of $Ce_2Sn_2O_7$, $J_{//}$ refers to the coupling between octupolar components of the pseudo-spins[39] – a possibility that was predicted by theory[13,40] and later observed experimentally[12]. The asymmetry observed in the data indicates $J_{\pm} < 0$, as confirmed by the fit (blue curve in Fig. 4**a**), because flipping the sign of $J_{\pm}$ in the spinon hopping model would otherwise invert the shape of the spectrum along the energy axis[14]. The data thus suggests[12] that $Ce_2Sn_2O_7$ stabilizes the $\pi$-flux phase of QSI – the symmetry enriched state occupying a large portion of the QSI phase space[41], as also argued in $Ce_2Zr_2O_7$ (refs. 42-43). In the $\pi$-flux phase, translational symmetry fractionalizes[32-33,19], so that the spinons acquire a finite Aharonov-Bohm phase after transporting them around any hexagonal plaquette, leading to an enhanced periodicity of the two-spinon density of states in the Brillouin zone[16-17]. It is noteworthy that the fitted value of $J_{//} = 48$ μeV $\approx 0.6$K is also in excellent agreement with a recent measurement of the specific heat of $Ce_2Sn_2O_7$ that shows a broad peak centered at $T \approx 0.12$ K (ref. 47), as quantum Monte Carlo simulations also predict this feature at $T \sim 0.2\,J_{//}$ (ref. 31).

A widely used theoretical framework to study QSI is gauge mean-field theory, where spinons hop on the centers of tetrahedra (c.f. Fig. 1) while interacting with the emergent U(1) gauge field[44,32]. A recent extension of this theory allows for the classification of symmetry fractionalization[16], predicting clear spectroscopic signatures for the $\pi$-flux phase[17]. The spinon dispersion is expected to be composed of two bands that are mostly flat, leading to three peaks in the two-spinon density of states, with energy separations proportional to $J_{\pm}/J_{//}$ and intensity ratios reproducing an overall asymmetric spectrum. We use these results[16-17] to fit $\chi''(E)$ including a phenomenological line broadening accounting for finite spinon lifetime and thermal fluctuations (black curve in Fig. 4**b**), giving $J_{//} = 69$ μeV and $J_{\pm} = -17$ μeV. This model provides an explanation for the scattering observed at $E > 80$ μeV that is not accounted for by the spinon hopping model where gauge fluxes are thermally activated and incoherent ($J_{ring} \ll T \ll J_{//}$). Although the intensity of the second peak is overestimated by the gauge mean-field theory at zero temperature, the level of agreement is remarkable given the nature of this model, i.e. the qualitative observation of peaks in the data and reproducing their positions in energy is significant. At finite temperatures $T \approx J_{ring}$, when gauge fluxes just start to freeze and become coherent, we expect that thermal fluctuations renormalize the relative intensities of the three peaks. The exchange constants translate into $J_{ring} = 12.4$ μeV, which indicates that our experiments at $T \approx 0.17$ K $\approx 15$ μeV were indeed performed inside the correlated phase (see Extended data fig. 2), in a temperature regime where quantum coherence is not completely destroyed by thermal fluctuations. The physical existence of a second peak in the two-spinon density of states, at approximately the same energy as in the gauge mean-field theory, is also confirmed by recent numerical results using exact diagonalization[45] (grey curve in Fig. 4**b**). Another outcome of the GMFT fit is to provide an estimate for the size of the two-spinon gap, which for the fitted $J_{\pm}/J_{//}$ amounts to $2\Delta \approx 33\ \mu\text{eV}$. This is higher than the onset visible in the data, around $18\ \mu\text{eV}$ – a difference that we attribute to line broadening effects.

For completeness, we also compare our $\chi''(E)$ data to the theoretical DOS considering the QED effects of spinons (electric charges) propagating on a coherent QSI background (photons)[15]. In this case, the most significant consequence of the Coulomb interaction is an abrupt step-function onset of spinon production at small momenta, which is known as the Sommerfeld enhancement[15]. In the emergent QED, depending on the exchange parameters, spinons can propagate much faster than the photons. This effectively leads to a broadening of the threshold at larger momenta, because spinons start to emit diffuse Cerenkov radiation[15]. The corresponding analytical model applies in the long-wavelength limit and thus can only be compared with our data at the low-energy onset of the spinon band. Therefore, we fit the analytical QED model to our HR backscattering data as shown with the dashed red lines in Fig. 4. The exchange parameters obtained from the spinon hopping model[14] or gauge mean field theory[16-17] can be converted to predefined parameters in the QED model – namely the ring exchange, spinon mass and speed of light. The fine-structure constant of the emergent QED – a dimensionless value characterizing how strongly light and matter couple, was fixed to $\alpha = 0.08$ based on numerical estimates for QSI[46]. After integrating the analytical model over the experimental window of momentum transfers, the calculated DOS matches the experiment well with a spinon gap $2\Delta \approx 18\ \mu\text{eV}$.

Finely resolving the spectrum of continuous excitations in a candidate QSI material opens the door to benchmarking important theory predictions on this unique quantum mechanical ground state. The agreement with the DOS expected for QSI is significant for several reasons. The characteristic features observed in the data – gap, threshold, main peak and asymmetry, are signatures of the strong interaction of fractional excitations with the emergent gauge field. The presence of a continuum with three approximately equally spaced peaks of decreasing intensity is a distinctive signature of the $\pi$-flux QSI that, as far as we know, is not predicted for any other phase on the pyrochlore lattice, implying that the experiment probes a specific signature of fractionalization[17]. For example, when quantum coherence of the QSI ground state is lost, the monopoles effectively hop on an ensemble of classical spin ice states background. Such spectrum, calculated analytically and numerically, shows only one peak with a long tail[14]. Thus, observing peaks in the spinon spectrum is evidence for the quantum coherence of the underlying QSI ground state. Moreover, we extract the exchange parameters of a QSL using a microscopic probe, directly from the spin liquid ground state excitation spectrum. This contrasts with the method of inferring exchange parameters from a related field-induced ordered phase. Recent numerical results have established how the emergent QED compares with that of our universe through estimates of its fine-structure constant[46]. It is predicted that the alternative vacuum of this condensed matter system is drastically different, with phenomena arising from strong light-matter interactions[15,46]. Our data cannot be used to directly determine the fine-structure constant, and drawing conclusions from a fit to the analytical QED model[15] would require experiments performed at much lower temperature. Momentum-resolved experiments on single-crystal samples would certainly further our understanding, however, directly fitting QED parameters from such data may require resolutions in both energy and momentum space that are far beyond current spectroscopic techniques. Using samples of $Ce_2Sn_2O_7$ prepared by a different chemical route but of equally high quality, a recent report interpreted the observation of dipolar spin ice scattering as a proximate state to an all-in-all-out ordered ground state[47]. Alternatively, this result could reflect a tiny sample-dependent change of the dipole-octupole coupling ($J_{xz}$ in the dipole-octupole Hamiltonian[40,13]), leading to a drastic change in neutron scattering while maintaining the dominant octupolar exchange responsible for the ice manifold ($J_{xx}$ or $J_{yy}$). Together with the fact that different ratios of exchange interactions were found in the three known cerium pyrochlore materials[42-43,48], this may suggest a high degree of tunability of the emergent QED, perhaps opening the door to its experimental control.

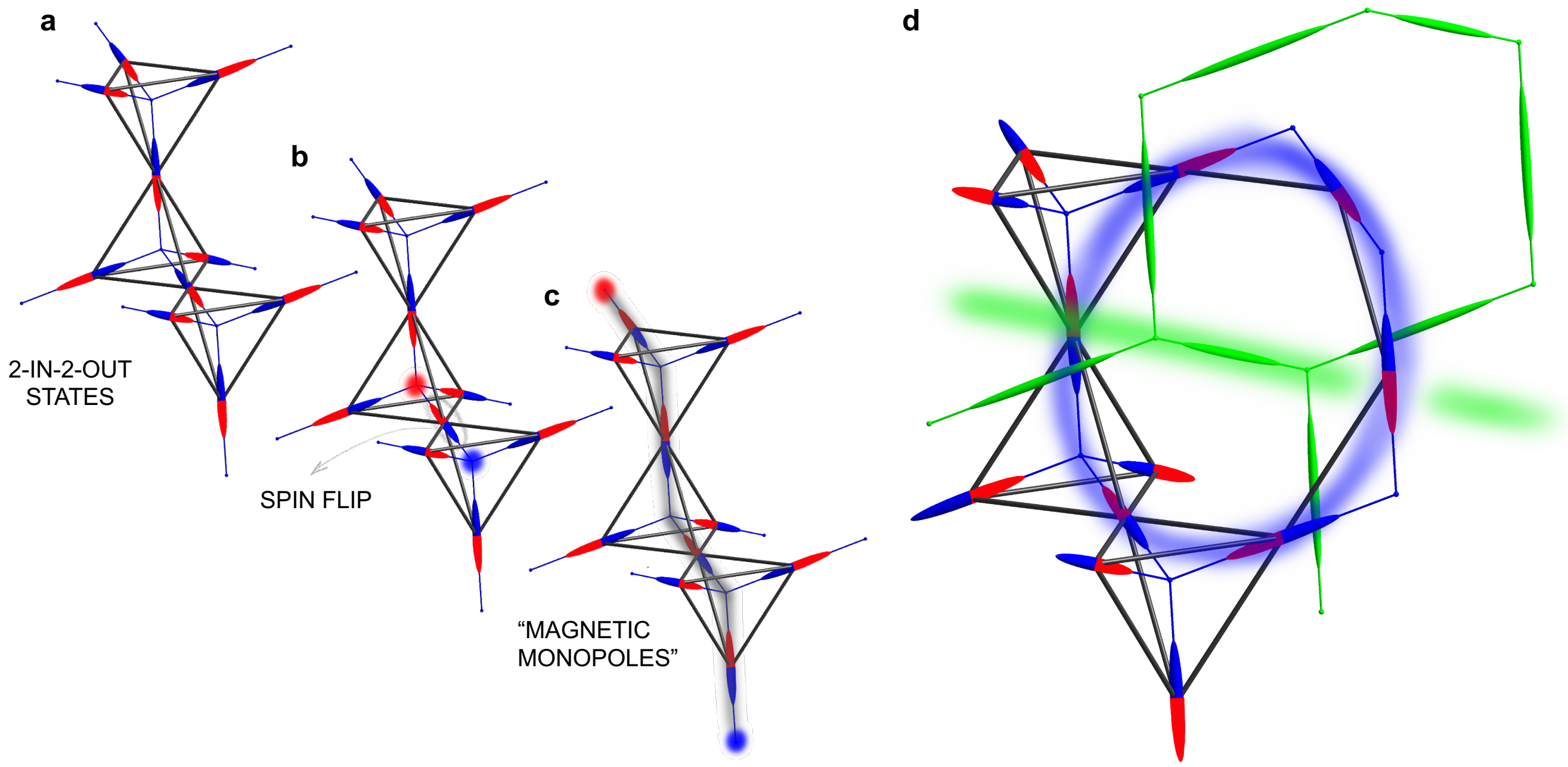

**Figure 1 | Correlations and excitations in quantum spin ices.** The '2-in-2-out' ice configurations found in spin ices (**a**), as well as the creation (**b**) and propagation (**c**) of 'magnetic monopole' quasiparticles. The ellipses represent uniaxial magnetic moments, with blue and red poles, defining magnetic flux variables that live on a diamond lattice (blue lines). In $Ce_2Sn_2O_7$ the ice rule applies similarly on objects of a more complex magnetization density (magnetic octupoles)[12,13,40]. In classical spin ice, thermal fluctuations create spin flips leading to fractional magnetic charges propagating through the sample (blue and red spheres)[27]. In a quantum spin ice (QSI)[5–6], the corresponding fractional gapped excitations (spinons) execute quantum coherent motion. The dominant tunneling process in QSI occurs on hexagonal plaquettes highlighted by the blue loop on panel **d**. This quantum dynamics is encoded by the fluctuation of electric flux variables living on a second diamond lattice (drawn in green) interpenetrating the first one. In this emergent quantum electrodynamics, transverse fluctuations of the dual gauge field are gapless 'magnetic photon' excitations[5–6].

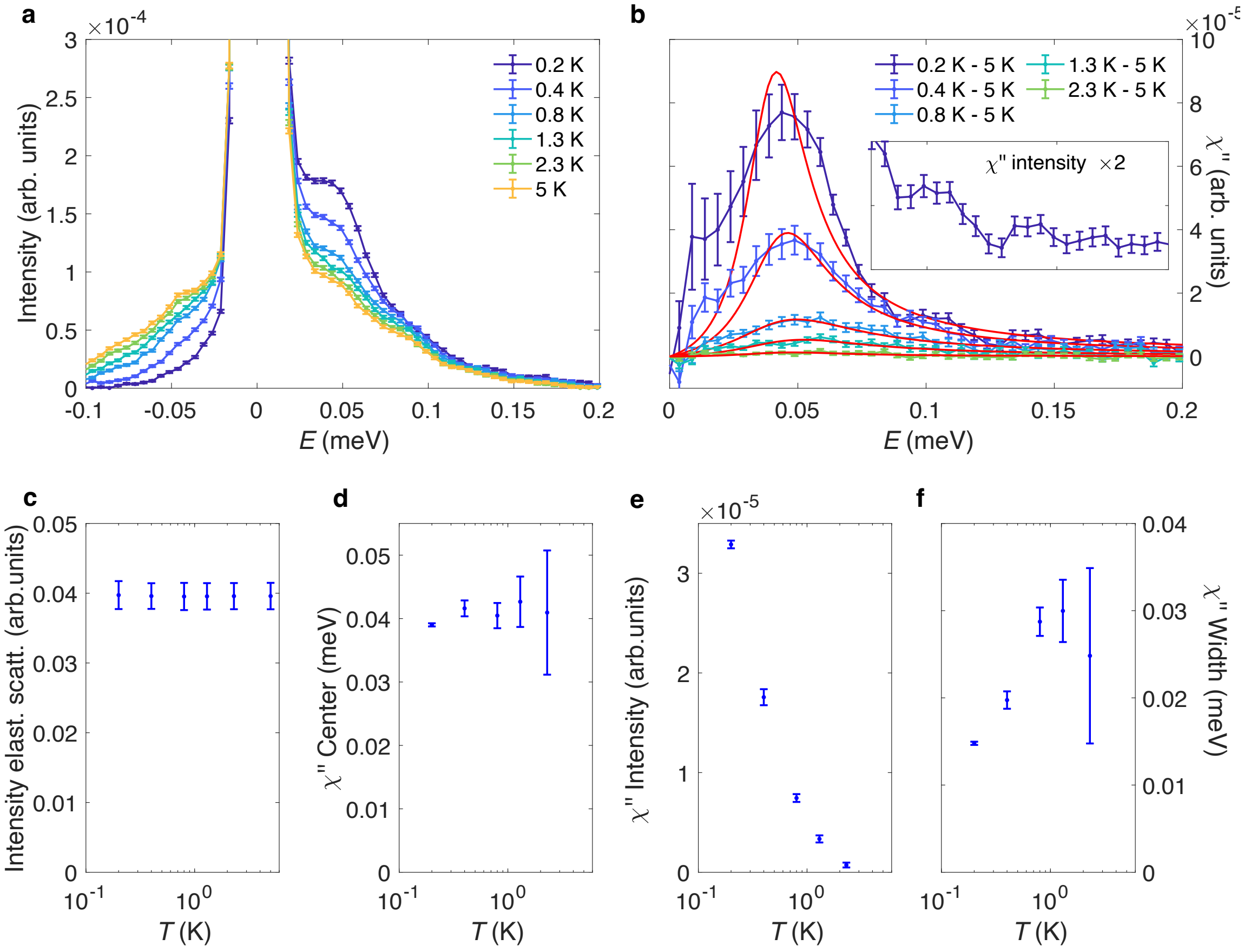

**Figure 2 | Temperature evolution of spin excitations in $Ce_2Sn_2O_7$. a**, Inelastic neutron scattering data measured at the time-of-flight spectrometer IN5 using an incident wavelength of 10 Å, providing an energy resolution of 11 μeV. The spectra were collected at various temperatures indicated in the plot, integrated on a range of momentum transfers $|\boldsymbol{Q}|$ from 0.3 to 1.1 $Å^{-1}$ and corrected for instrumental background, resulting in the experimental data points with error bars corresponding to ±1 standard error. **b**, Imaginary part of the dynamic spin susceptibility $\chi''(E)$ (data points with error bars corresponding to ±1 standard error) extracted from the data shown in panel **a**, as described in the main text. The red lines represent phenomenological Lorentzian fits of the data, allowing to numerically track the temperature dependence of the spin excitations. The fit function is defined as $\chi''(E) = \frac{S_f \gamma E}{(E-\delta)^2+\gamma^2}$, with $S_f$ a global scale factor, $\gamma$ the Lorentzian width and $\delta$ its center. Panel **c** shows the temperature evolution of the scattering at the elastic line, with error bars corresponding to ±1 standard error, indicating that the magnetic scattering in the accessible $|\boldsymbol{Q}|$ range is essentially inelastic. Panels **d**, **e** and **f** present the temperature dependence of the center, intensity and width of the Lorentzian fit to the $\chi''(E)$ data, respectively, with error bars corresponding to ±1 standard error.

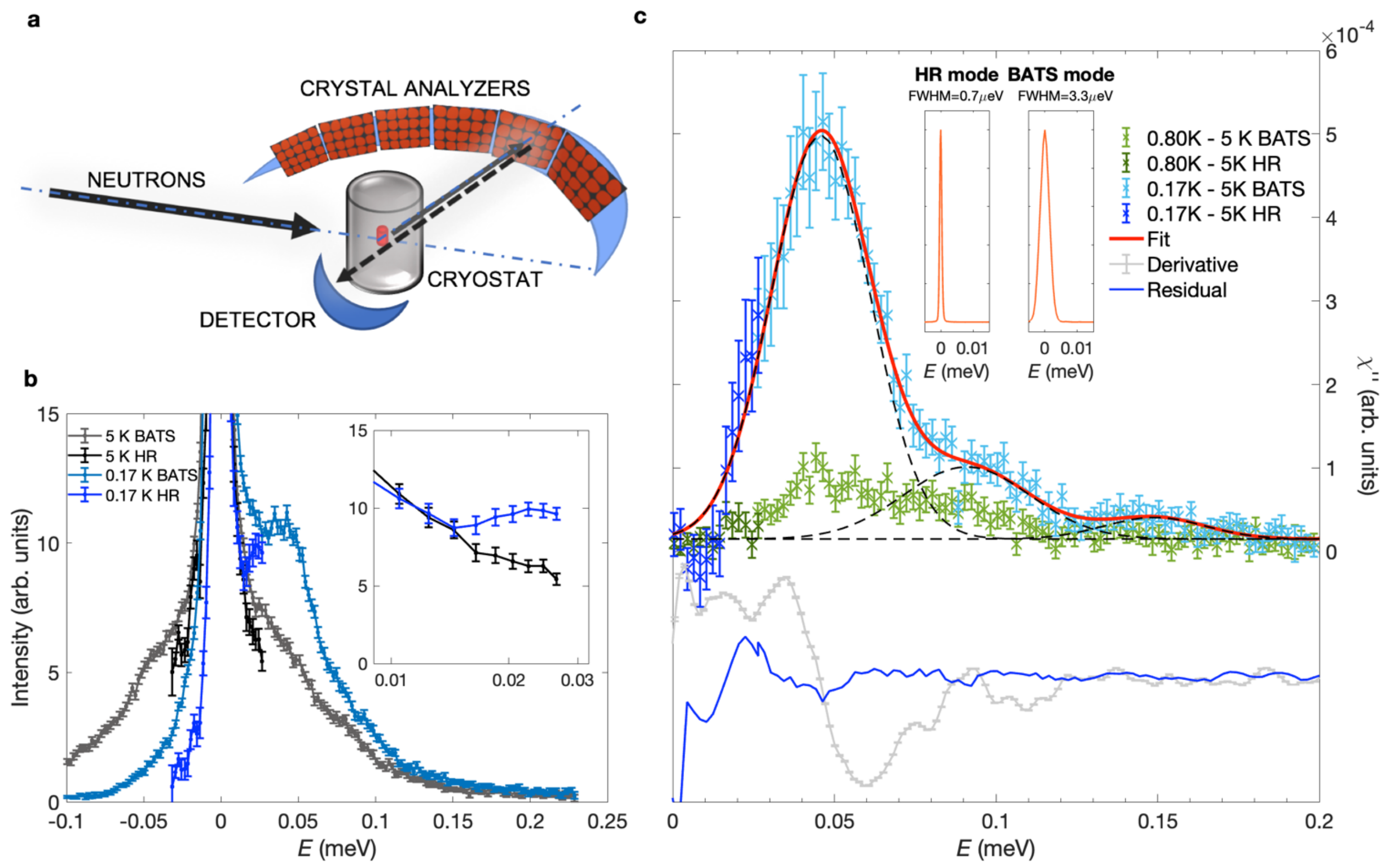

**Figure 3 | High-resolution neutron spectroscopy of fractional excitations in $Ce_2Sn_2O_7$. a**, Sketch of the neutron backscattering technique. Neutrons are first scattered by the sample towards crystal analyzers – a component that discriminates their energy with a very high resolution, and then backscattered towards a detector[10]. **b**, Comparison of the $Ce_2Sn_2O_7$ spectra collected at 0.17 K and 5 K using the IN16B instrument (Institut Laue–Langevin, Grenoble) in 'backscattering and time-of-flight spectrometer' (BATS)[36] and 'high-resolution' (HR)[50] modes. The spectra (data points with error bars corresponding to ±1 standard error) were integrated on an identical range of momentum transfers |***Q***| (0.4 to 1.7 $Å^{-1}$) and rescaled on the basis of their respective elastic line intensities, effectively correcting any discrepancies between the two configurations. The inset shows a zoom into the HR data, focusing on the threshold part of the spectra and showing the clear rise of the continuum on top of the remaining paramagnetic quasi-elastic signal. The latter is attributed to fluctuations of the dipole components of the pseudo-spin at finite temperatures. **c**, Superposition of the imaginary part of the dynamical spin susceptibility, $\chi''(E)$ (data points with error bars corresponding to ±1 standard error), extracted from the continuation of HR (dark colored symbols) and BATS (light colored symbols) experiments at 0.17 K (blue shades) and 0.8 K (green shades). The continuous red line is a fit using a constant background and three Gaussian peaks individually shown as dashed black curves. The centers, intensities and widths are unconstrained, confirming the significance of each contribution to the spectrum. The continuous blue line and the grey points are the residual of the fit and the derivative of the experimental data, respectively, both shifted by -1.5 arbitrary units for clarity. The two insets in panel **c** show the energy resolution provided by each instrument configuration, on the same energy scale as the main panel.

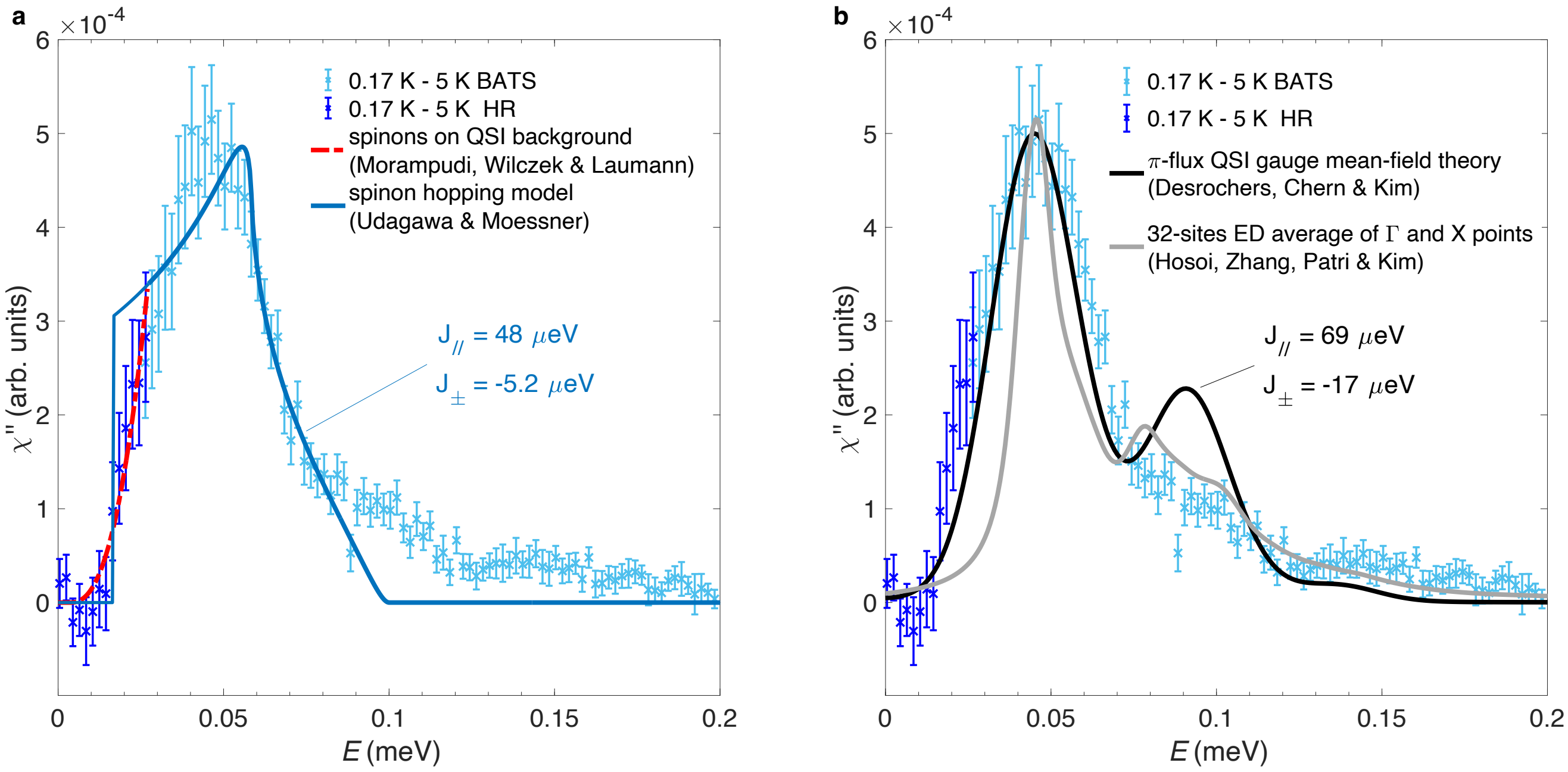

**Figure 4 | Comparison of the dynamical spin susceptibility with models of spinon dynamics for the $\pi$-flux phase of quantum spin ice.** In **a**, the continuous blue line is a fit of the combined HR and BATS data on their full energy window at 0.17 K, using the analytical model of Udagawa & Moessner for the quantum dynamics of spinons hopping on a lattice and considering a classical spin ice background[14]. Similarly, in **b**, we show the best fit using the gauge mean field theory of quantum spin ice revised by Desrochers, Chern & Kim[16-17]. In both these fits, the adjustment variables are exchange parameters $J_{//}$ and $J_{\pm}$ whose fitted values are indicated in the respective panels, corresponding to $J_{\pm}/J_{//} = -0.1083$ (**a**) and $J_{\pm}/J_{//} = -0.2464$ (**b**). The fit using the gauge mean-field theory incorporates a peak broadening (standard deviation $\sigma = 11.2\ \mu\text{eV}$). In panel **b**, we also compare the fit with available results of numerical calculations for $J_{\pm}/J_{//} = -0.1875$ using exact diagonalization on 32 sites (Hosoi, Zhang, Patri & Kim[45]). The corresponding curve (solid grey line) is the average of results at the $\Gamma$ and X points of the Brillouin zone after setting the energy scale of $J_{//}$ to the value determined from the fit of the gauge mean field theory. For completeness, we also compare the onset of the continuum to the analytical model of Morampudi, Wilczek & Laumann considering a QSI background (red dashed lines in panel **a**), i.e., including photons, which effectively broadens the threshold for our experimental |***Q***| window due to the emission of Cerenkov radiation[15]. These QED effects are neglected in the other models, while the model of Morampudi *et al.* neglects the lattice and therefore can only be used to compare with data at the lower edge of the continuum. We used a numerical estimate for the fine-structure constant, $\alpha = 0.08$ (ref. 46), and other QED parameters obtained from the conversion of the exchange parameters $J_{//}$ and $J_{\pm}$ – see Methods. The red dashed line in panels **a** is a fit with only one free parameter, the two-spinon gap $2\Delta \approx 18\ \mu\text{eV}$.

**Acknowledgements**

This work is based on experiments performed at the Institut Laue–Langevin, France. We thank Xavier Tonon, Eric Bourgeat-Lami and the whole team for Advanced Neutron Environments for their dedicated work running the dilution refrigerators at the Institut Laue–Langevin. Tom Fennell is warmly acknowledged for his continuous support throughout this project and for a careful reading of the manuscript. We thank Nic Shannon for fruitful discussions. We acknowledge funding from the Swiss National Science Foundation (R.S. and V.P., Grant No. 200021_179150), the U.S. National Science Foundation Division of Materials Research under the award DMR-1917511 (H.Y. and A.H.N.), and the Natural Sciences and Engineering Research Council of Canada (F.D. and YB.K.).

**Author contributions**

Project and experiments were designed by R.S. Sample preparation and characterization were performed by R.S. and V.P. Neutron scattering experiments were carried out by V.P., E.L., S.P. and R.S. with O.J. and M.A. as local contacts. Experimental data were analysed by V.P., E.L., S.P. and R.S. Calculations were performed by H.Y., F.D., YB.K. and A.H.N. The paper was written by R.S. with feedback from all authors.

**Competing financial interests**

The authors declare no competing financial interests.

**Methods**

**Sample preparation**

The sample used during this work is a large polycrystalline sample of $Ce_2Sn_2O_7$, which was also used in previous studies[11–12]. The solid-state synthesis product results from an oxido-reductive reaction where metallic tin is employed to reduce cerium to the trivalent state. We note that evaporation of tin or tin oxide is not an issue given their vapor pressures at the temperature of this synthesis ($\approx 10^{-7}$ bar at 1000 Celsius), especially with materials slowly pre-reacted at lower temperature[12]. The sample was thoroughly investigated using diffraction techniques as well as thermogravimetry, confirming the high purity of the sample and its stoichiometry $Ce_2Sn_2O_{7.00\pm0.01}$, which indicates that all cerium ions are magnetic and experience the same crystal-electric field environment. Fits of neutron pair distribution function data further indicate the absence of disorder or local distortions of the crystal structure[12].

**Neutron scattering experiments**

The time-of-flight inelastic neutron scattering experiment was performed with the IN5 time-of-flight spectrometer at the Institut Laue-Langevin, Grenoble, France[49]. The $Ce_2Sn_2O_7$ powder was pressed into pellets and inserted inside a copper can (10 mm diameter, filled on a height of ~5 cm), which was then sealed, pressurized using 10 bars of helium gas at room temperature, and mounted below the mixing chamber of a dilution refrigerator. The helium overpressure was required in order to maximize cooling efficiency. An incident wavelength of 10 Å was used, providing an energy resolution of about 11 μeV at the elastic line. The cooling of the sample was monitored by following the evolution of the inelastic contribution to the signal. The temperature of the sample was estimated by fitting the spectrum using a phenomenological background function, a Gaussian centered at zero energy transfer and two gapped Lorentzian multiplied by $1/(1-\exp\left(-\frac{E}{k_B T}\right))$ to account for temperature effects. The datasets were recorded in such a way to obtain similar statistics at each temperature. Calibration scans (vanadium and empty copper can) were used to properly reduce the data using Mantid[50] routines, resulting in six pre-processed datafiles containing the $\|\vec{Q}\|$-integrated spectra (0.3 $Å^{-1}<\|\vec{Q}\|<1.1$ $Å^{-1}$).

Neutron backscattering spectroscopy was performed on IN16B at the Institut Laue-Langevin[36,51]. The sample and sample preparation were identical to the IN5 experiment described above. In a first step, we used the BATS mode available at IN16B (ref. 36) in order to cover the full bandwidth of excitations in $Ce_2Sn_2O_7$. Two instrument configurations, denoted as lr4 and lr6, were used and correspond to low repetition rates with 8° and 11° slits in the pulse choppers, providing a resolution of 4 μeV and 6 μeV respectively. The lr6 allowed to efficiently measure spectra at intermediate temperatures, benefiting from a more intense beam at the expense of a slightly coarser resolution with respect to lr4. The thermalization of the sample was monitored as described for the IN5 experiment. Spectra were recorded at 0.17 K, 0.8 K and 5 K using both lr4 and lr6, with additional measurements at 0.4 K and 1.2 K with the lr6 set-up. Data for a vanadium standard, empty copper can and empty dilution were also recorded and used in the reduction routines using Mantid[50]. The spectra were integrated over the same $\|\vec{Q}\|$ window ranging from 0.4 $Å^{-1}$ to 1.7 $Å^{-1}$. The resulting data can be seen in Fig. 3**b** and Fig. S1**a**, for lr4 and lr6 respectively. The imaginary part of the dynamical spin susceptibility was computed following the same method as described above and the results are shown in Fig. 3**c** (as well as Fig. 4**a** and 4**b**) and Fig. S1**b**, for lr4 and lr6, respectively.

A second experiment was carried out on IN16B, in order to better investigate the lower part of the energy spectrum. The High-Resolution (HR) mode of the instrument was used, which has a lower flux compared to the previously mentioned BATS mode, and a resolution at the elastic line of about 0.7 μeV. We have used a specialized high signal-to-noise ratio setup of the IN16B spectrometer previously reported[51]. The same powder sample was again used but this time was loaded in a copper can with annular geometry (outer 15 mm, inner 10 mm). The reason for such a choice was the reduction of the neutron absorption by the sample, which in this geometry, plays a more important role. The sample was cooled down to an estimated base temperature of approximatively 0.17 K. Data were recorded at three different temperatures, 0.17 K, 0.8 K and 5 K with similar statistics, allowing to track the signal's behavior and a direct comparison with previous experiments. The data were reduced via Mantid[50] routines, using carefully measured calibration scans (vanadium sheets, empty annular copper can and empty dilution refrigerator). The resulting spectra were then integrated over a $\|\vec{Q}\|$ window ranging from 0.4 $Å^{-1}$ to 1.7 $Å^{-1}$. The final spectra can be seen in Fig. 3**b**. The imaginary part of the dynamical spin susceptibility was computed following the same method as described above and is plotted in Fig. 3**c** as well as in Fig. 4**a** and 4**b**. In order to get a meaningful comparison of the BATS and HR data, the lr4 HR spectra were subject to a minor rescaling, based on the relative intensities at the elastic line, thus compensating for any discrepancies between the two instrument modes.

**Fitting of the experimental data to model calculations**

We consider a Hamiltonian where the transverse exchange parameter $J_{\pm}$ introduces quantum fluctuations to a classical spin ice manifold obtained from a dominant nearest-neighbor interaction $J_{\parallel}$:

$\mathcal{H}_{QSI} = \mathcal{H}_{CSI} + \mathcal{H}_{transverse} = \sum_{(i,j)} J_{\parallel} S_i^y S_j^y - J_{\pm}(S_i^+ S_j^- + S_i^- S_j^+)$ . Here $S^y$ corresponds to the octupolar component of the 'dipole-octupole' pseudo-spin[40], stabilizing an octupole ice manifold in $Ce_2Sn_2O_7$ (ref. 12).

We first used the results of Udagawa and Moessner[14] to compare with our data. They found that the two-spinon density of state (DOS) can be well approximated by the following exact result $\rho_{\mathrm{HC}}^{(2)}(\omega) = \int \mathrm{d}\epsilon \rho_{\mathrm{HC}}^{(1)}(\omega - \epsilon) \times \rho_{\mathrm{HC}}^{(1)}(\epsilon)$ where $\rho_{\mathrm{HC}}^{(1)}(\epsilon) = \frac{3}{2\pi}\frac{1}{6-\epsilon}\sqrt{\frac{5-\epsilon}{3+\epsilon}}$ is the single spinon DOS with $\epsilon = (\omega - J_{\parallel})$ ) in arbitrary units. We compare $\rho_{\mathrm{HC}}^{(2)}(\omega)$ directly with our experimental data. We vary an overall scale factor and the parameters $J_{\parallel}$ and $J_{\pm}$ (after converting them into the meV units), to minimize the least-mean square difference between the theory and all the experimental data (BATS Ir4, BATS Ir6 and HR): $C_1 = \sum_{\omega \text{ in exp.}} \left(I_{\mathrm{exp}}(\omega) - a \times \rho_{\mathrm{HC}}^{(2)}(\omega)\right)^2$. Here, the parameters $J_{\parallel}$, $J_{\pm}$ are inside the definition of $\rho_{\mathrm{HC}}^{(2)}(\omega)$ but we did not write them out explicitly to lighten the notation. $J_{\parallel}$ is used in defining $\epsilon = \omega - J_{\parallel}$ in the single spinon DOS, and $J_{\pm}$ is determined when converting the unit of $\epsilon$ to meV. Here, the parameter $a$ is the overall scaling factor that we also fit. We found that the square sum is minimized by the following parameters $J_{\parallel} = 48\ \mu\text{eV}$, $J_{\pm} = -5.2\ \mu\text{eV}$ and $J_{\text{ring}} \equiv \frac{12J_{\pm}^3}{J_{\parallel}^2} = 0.73\ \mu\text{eV}$.

Second, we fit the experimental $\chi''(E)$ data to the gauge mean-field theory results of Desrochers, Chern and Kim[16-17]. We used the positions identified from the data as starting values for the centers of the three peaks expected for the $\pi$-flux phase of quantum spin ice. The goodness of fit measures are defined as

$$\chi_{Direct}^2 = \sum_n \frac{(I^{Exp.}(E_n) - I^{Theo.}(E_n))^2}{(\Delta I^{Exp.}(E_n))^2}$$

and

$\chi_{Peak}^2 = \sqrt{\sum_{i=1}^{3}(E_i^{Theo.} - E_i^{Exp.})^2 / E_i^{Exp.}}$.

We normalize both $\chi_{Direct}^2$ and $\chi_{Peak}^2$ before taking the weighted sum. The final goodness of fit is $\chi^2 = \alpha\, \chi_{Direct}^2 + (1-\alpha)\, \chi_{Peak}^2$ with $\alpha = 0.6$. We found that $\chi^2$ is minimized using $J_{\parallel} = 69\ \mu\text{eV}$, $J_{\pm} = -17\ \mu\text{eV}$ and $J_{\text{ring}} \equiv \frac{12J_{\pm}^3}{J_{\parallel}^2} = 12.4\ \mu\text{eV}$.

Finally, we used the above extracted exchange parameters from both the analytical model of Udagawa and Moessner[14] and the extended gauge mean-field theory of Desrochers, Chern and Kim[16-17], and applied these to the analytical model of Morampudi, Wilczek & Laumann[15], which determined neutron scattering as

$S(q,\omega,\Delta) = \frac{m^{3/2}\sqrt{2\pi R}}{1-\exp\left(-\sqrt{\frac{2\pi R}{\omega - 2\Delta - q^2/4m}}\right)}\theta(\omega - 2\Delta - q^2/4m)$, where $R = \frac{1}{4}mc^2\alpha^2\left(1 - \frac{q^2}{4m^2c^2}\right)^2$.

Most parameters in this model are determined by the spin exchange parameters: the loop flipping term coefficient $g = 12\frac{J_\pm^3}{J_\parallel^2}$, the spinon mass $m = \frac{1}{4J_\pm a_0^2}$, and the speed of light $c = \xi g a_0$. In addition, there are three constants independent of the value of $J_\parallel, J_\pm$, which are either known experimentally – the lattice constant $a_0 = 10.6 \times 10^{-10}$ m, or taken from numerical estimates[47] – the emergent fine-structure constant $\alpha = 0.08$ and the O(1) constant $\xi = 0.51$. Therefore, a fit to experimental data using this QED model has only two free parameters. The first free parameter is the overall scale of the DOS, while the second one is the spinon gap Δ. Although Morampudi *et al.*[15] take the gap to be $\Delta \sim J_\parallel/2 - 12J_\pm$, this value turns out to be negative from our fitting result. We hence take $\Delta$ to be a free parameter when fitting the experimental neutron intensity to the theory. Since the work of Morampudi *et al.*[15] is applicable in the long wavelength limit and does not consider the short wavelength effects of the pyrochlore lattice, it can only be used to compare with the low-energy end of the neutron scattering. In order to compare the model with the experimental data, we integrated over $q$ to obtain the (local) density of states distribution $\tilde{S}(\omega, \Delta) = \int \mathrm{d}q S(q, \omega, \Delta)$ and minimized the following quantity: $C_2 = \sum_{\omega \text{ in exp.}} \left(I_{\text{exp}}(\omega) - a \times \tilde{S}(\omega, \Delta)\right)^2$. Here, $a$ is the overall scaling factor and we use the low-energy HR dataset, which covers energy transfers up to 26.5 μeV. The best fit we found results in $2\Delta = 18$ μeV using the exchange parameters deduced from the analytical model of Udagawa and Moessner[14]. A similar curve can be obtained using the exchange parameters deduced from our fit to the extended gauge mean-field theory of Desrochers, Chern and Kim[16-17].

**Data availability**

The data that support the plots within this paper and other findings of this study are available from the corresponding authors upon reasonable request. The datasets for the time-of-flight neutron spectroscopy experiments on IN5 and for the backscattering neutron spectroscopy experiments on IN16B are available from the Institute Laue-Langevin data portal[52–55].

**References (methods)**

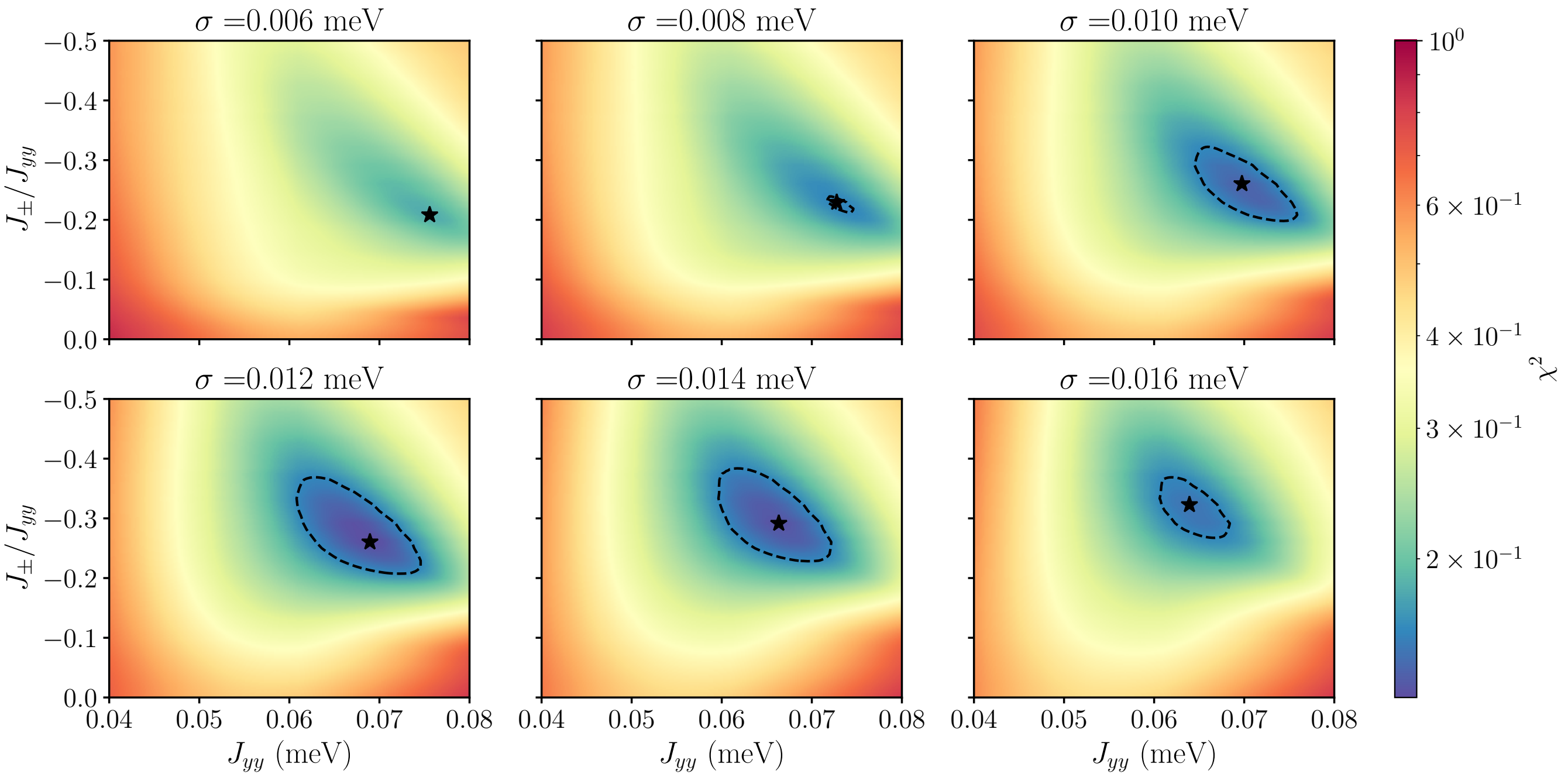

**Extended data Fig. 1 | Goodness of fit for the adjustment of the exchange parameters of the gauge mean field theory to the $\chi''(E)$ data.** The color maps show the goodness of fit defined in the Methods section, as a function of the dominant and transverse exchange parameters, for different values of the phenomenological peak broadening $\sigma$ (standard deviation). Here, $J_{yy}$ is an octupolar exchange interaction designating the dominant nearest-neighbor interaction $J_{\parallel}$. The fitting procedure is described in the Methods section and the final fit is presented in Fig. 4**b** (black curve). The hatched region denotes parameters for which the goodness of fit is below a threshold arbitrarily fixed at 0.17.

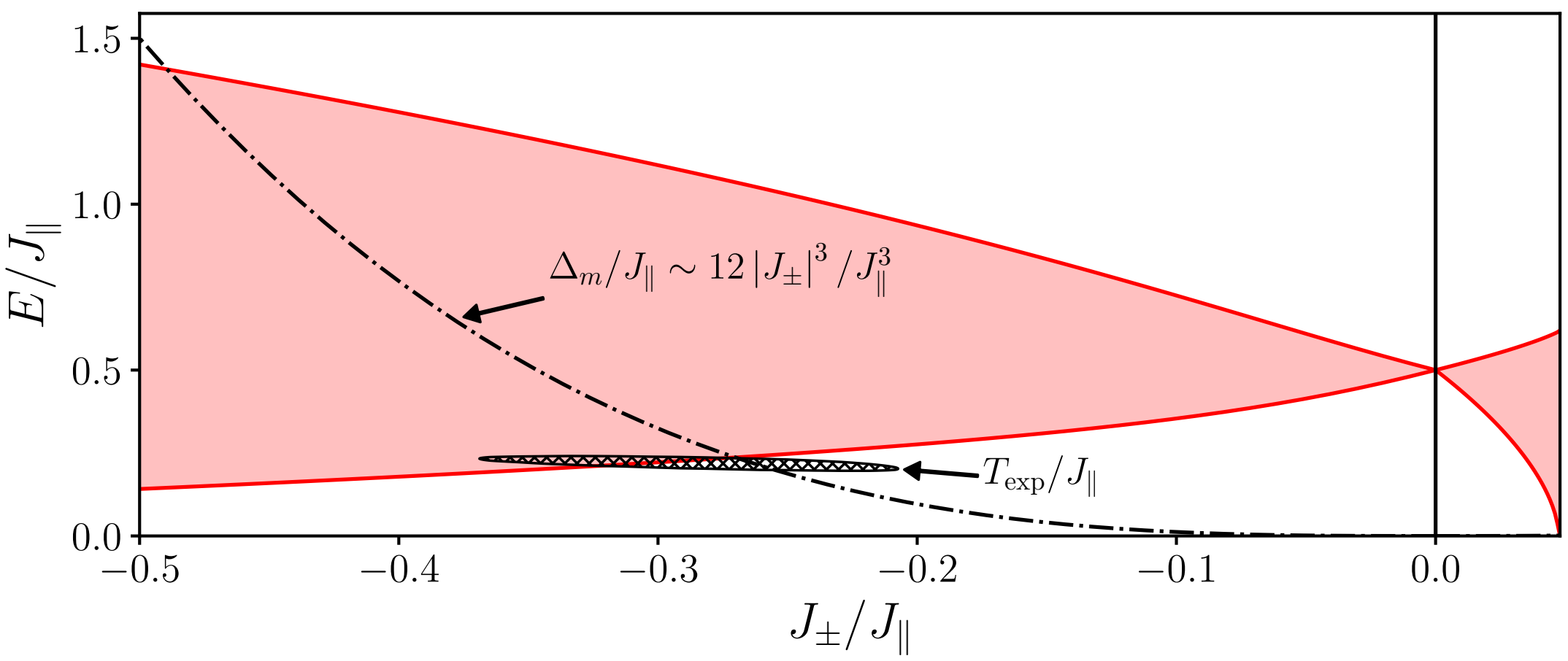

**Extended data Fig. 2 | Spinon dispersion's minimum/maximum** (red lines) **and the approximate vison gap** (black line) **as a function of $J_{\pm}/J_{\parallel}$** (ref. 18)**.** The plot highlights that the energy scale of the spinon and vison become comparable for the $\pi$-flux phase. The hatched region corresponds to our experiments based on i) the sample temperature evaluated from the detailed balance between neutron energy loss and gain sides and ii) the estimate of $J_{\pm}/J_{\parallel}$ corresponding to the best agreements between our data and the GMFT results (c.f. region of the goodness of fit encircled with a black dashed line, in Extended data Fig. 1). Assuming $J_{\parallel} \sim 0.069$ meV = 0.8 K, the sample temperature of our neutron backscattering experiment (T = 0.17 K) is inside the spin ice regime.